\documentclass[prl,showpacs,showkeys,preprintnumbers,amsmath,amssymb,superscriptaddress,twocolumn]{revtex4-1}
\usepackage[english]{babel}
\usepackage{makeidx} 
\usepackage{graphicx} 
\usepackage{dcolumn} 
\usepackage{array} 
\usepackage{amssymb} 
\usepackage{amsmath}
\usepackage{textcomp}
\usepackage{multirow}
\usepackage{subfigure}
\usepackage{eucal}
\usepackage{mathrsfs}
\usepackage[all]{xy}

\usepackage{color}

\usepackage{float} 
\usepackage{amsmath} 
\usepackage{amsfonts} 
\usepackage{bm} 

\begin{document}

\title{Carboxylate based molecular magnet: one path toward achieving stable quantum correlations at room temperature}

\author{C. Cruz\footnote{clebsonscruz@yahoo.com.br}} \affiliation{Instituto de F\'{i}sica, Universidade Federal Fluminense, Av. Gal. Milton Tavares de Souza s/n, 24210-346 Niter\'{o}i, Rio de Janeiro, Brazil.} 
\author{D. O. Soares-Pinto} \affiliation {Instituto de F\'{i}sica de S\~{a}o Carlos, Universidade de S\~{a}o Paulo, CP 369, 13560-970, S\~{a}o Carlos, SP, Brazil}
\author{P. Brand\~ao} \affiliation{CICECO, Universidade de Aveiro - 3810-205, Aveiro, Portugal, EU}
\author{A. M. dos Santos} \affiliation{Quantum Condensed Matter Division, Oak Ridge National Laboratory - Oak Ridge, TN 37831-6475, USA}
\author{M. S. Reis} \affiliation{Instituto de F\'{i}sica, Universidade Federal Fluminense, Av. Gal. Milton Tavares de Souza s/n, 24210-346 Niter\'{o}i, Rio de Janeiro, Brazil.}


\date{\today}

\begin{abstract}

The control of quantum correlations in solid state systems by means of material engineering is a broad avenue to be explored, since it makes possible steps toward the limits of quantum mechanics and the design of novel materials with applications on emerging quantum technologies. In this context, this Letter explores the potential of molecular magnets to be prototypes of materials for quantum information technology. More precisely, we engineered a material and from its geometric quantum discord we found significant quantum correlations up to 9540 K (even without entanglement); and, in addition, a pure singlet state occupied up to around $80$ K (above liquid nitrogen temperature). These results could only be achieved due to the carboxylate group promoting a metal-to-metal huge magnetic interaction. 
\end{abstract}
\maketitle


Quantum entanglement has received a considerable attention as a remarkable resource for quantum information processing\cite{nielsen,vedral6,horodecki}. In spite of that, it is fragile and can easily vanish due to the inevitable interaction of the system with its environment\cite{yu}; and due to this condition, it was thought that entanglement could only exist at low temperatures. However, recently, it has been shown that entanglement can also exist at higher temperatures, and can be detected through the measurement of some thermodynamic observables\cite{vedral5,diogo,duarte,duarte2,diogo3,souza2,mario2,vedral2,souza,vedral7,toth,ghosh,vedral8}.

Nevertheless, quantum entanglement does not encompass all quantum correlations in a system and recent studies have greatly expanded the notion of quantum correlations\cite{modi,zurek,vedral,vedral9,liu,ma,adesso,sarandy,luo,datta,sarandy2,vedral3}; and the measure of quantum excess of correlations has been named as \textit{quantum discord}\cite{zurek,vedral,vedral9}. In the last years, it was understood that quantum discord has an important role in many quantum information processing even without entanglement. Notably, this quantity can also detect quantum phase transitions \cite{pirandola,sarandy,werlang}.

Despite much effort by the scientific community, there are only a few results on the analytical expression of quantum discord; and only for a certain class of states an exact solution is known\cite{ma,luo,sarandy,adesso,datta,vedral4,terno}. This fact stimulated alternative measurements of quantum discord, theoretically and experimentally\cite{liu,adesso,vedral3,luo2,yuri,yuri2}. 
The recent demonstration that quantum discord can be measured by the thermodynamic properties of solids, such as magnetic susceptibility, internal energy \cite{yuri,yuri2,singh}, specific heat \cite{yuri,yuri2} and even neutron scattering data \cite{liu}, shows that quantum correlations can be related to significant macroscopic effects 
allowing the measurement and the control of quantum correlations in solid state systems by means of material engineering. Thus, the design of novel materials becomes an actual challenge to overcome.

In this direction, molecular magnets can be an excellent opportunity to achieve this goal as prototypes of materials for quantum information technology. They combine classical properties, found in any macroscopic magnet, with quantum ones, such as quantum interference and entanglement. The large gap between the ground state and the first-excited state found in some materials allows the existence of entangled quantum states at higher temperatures\cite{diogo,mario2,souza,vedral8}. 
The study of molecular magnets opens thus doors for new research toward to the limits of quantum mechanics; and several aspects are worthy of a thorough research, such as (i) how robust their quantum features are against temperature and (ii) how large systems can support these properties. These results would lead to promising applications in quantum technologies, specially devices based on quantum correlations.

In this context, we show in this Letter that carboxylate based compounds can support quantum correlations thousands of kelvins above room temperature. We report the crystal structure and magnetic susceptibility of a carboxylate based molecular magnet with formula $\left[ Cu_2(HCOO)_4(HCOOH)_2(piperazine)\right]$; from which the entropic and geometric quantum discord, based on the Schatten 1-norm\cite{ciccarello,obando}, are extensively explored. The analytical formulas for the quantum correlations are derived as a function of the temperature using the magnetic susceptibility of the compound and the analysis of the data suggests the existence of entanglement up to temperatures of $681$ K, while the measure of quantum discord reveals that this system keeps at the singlet ground state up to around $80$ K. In addition, quantum correlations would remain up to $9540$ K, thousands of kelvins above room temperature, even without entanglement. Thus, we obtain very stable quantum correlations up to $513$ K, the limit where this material can exist - hundreds of kelvins above room temperature \cite{paula}. Due to the material topology, these results represent the highest temperatures, reported to date, wherein quantum correlations can be supported.

\begin{figure}[htp]
\centering
\subfigure[]{\includegraphics[scale=0.26]{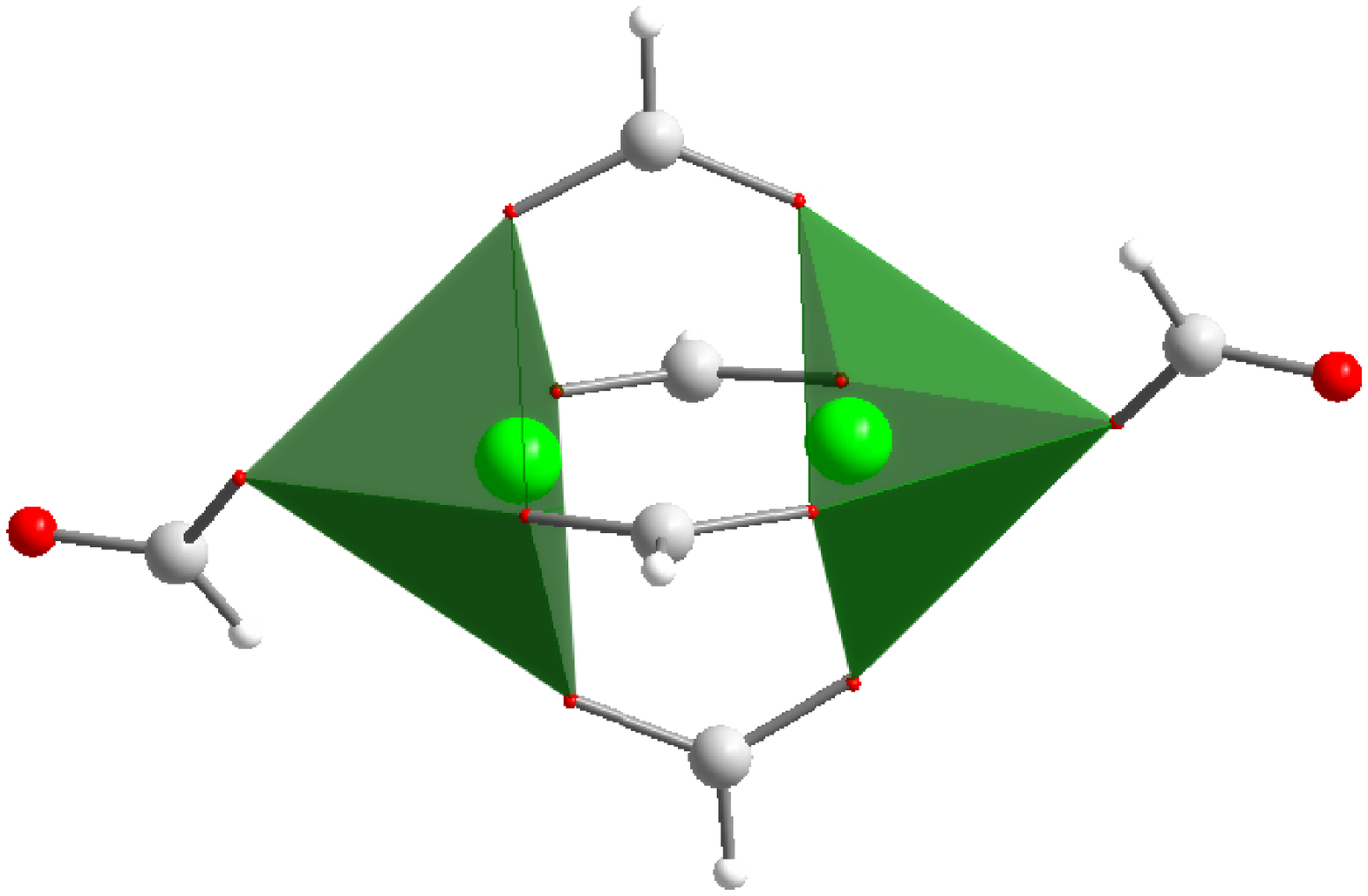}}
\subfigure[]{\includegraphics[scale=0.1]{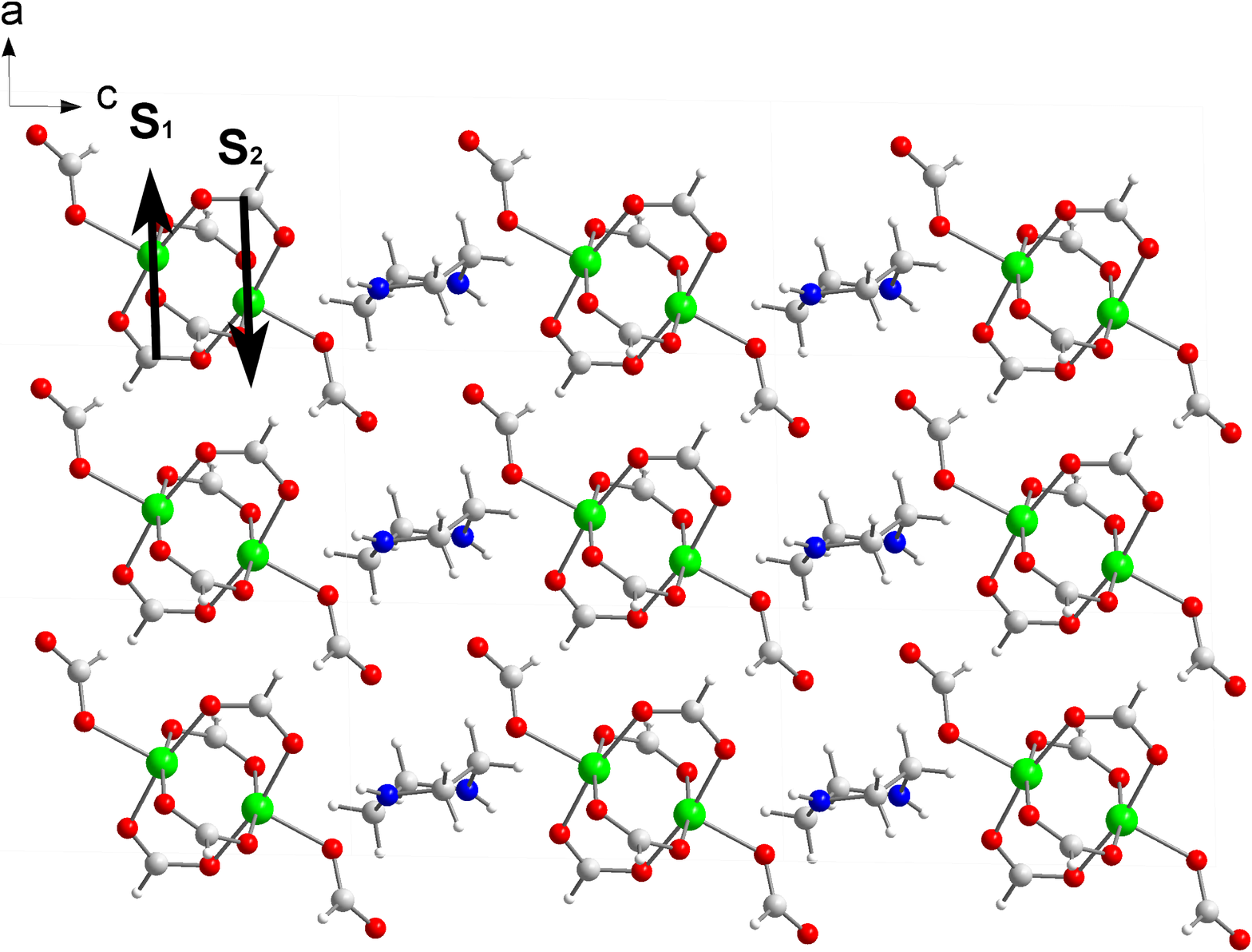}}\qquad
\caption{(color online) Crystal structure of $[Cu_2(\mu-HCOO)_4(HCOO)_2]piperazine$ compound with (a) local dimer polyhedron representation and (b) dimeric sheet view, where the arrows represent the reduced magnetic structure of the compound. Color scheme: Cu - green; O - €"red; N - €"blue.}
\label{fig:fig1}
\end{figure}

As pointed out before, to obtain such strongly entangled states at high temperature, we need to maximize the gap between the ground state and the first-exited state. Thus, we have designed materials with such features to enhance their quantum properties, and then study the entropic and geometric quantum discord.
The dinuclear copper(II) complex $[Cu_2(\mu-HCOO)_4(HCOO)_2]piperazine$ was the successful prototype synthesized, and its molecular structure is presented on figure\ref{fig:fig1}. 

Single-crystal structure analysis revealed that the compound is composed of a dimeric cupric tetraformate unit, with a short $Cu-Cu$ internuclear separation of $2.628(2)$ \AA \ and one piperazine molecule. 
The dimer is formed by opposing square pyramidal $CuO_5$ with a very small distortion - see figure \ref{fig:fig1}(a). The base oxygen atoms on the adjoining pyramids are part of the four-connecting carboxylate groups in a syn-syn conformation leading to a strong magnetic interaction between the dimers ions. The apical oxygen of the pyramid is connected, via $HCOO$ group ($Cu-O=2.12(1)$ \AA), to other carboxylate group. The crystal structure is stabilized by the intermolecular $N-HO$ hydrogen bonding with $N(10)\cdots O(7)= 2.722(2)$\AA \ and $N(10)\cdots O(8)^{ii}=2.745(2)$\AA \ ($ii=1+x, y, z$) forming an infinite two dimensional network.

From the magnetic point of view, the designed compound $[Cu_2(\mu-HCOO)_4(HCOO)_2]piperazine$ is of highest interest, since it shows a short interaction path between the metal centers. From their crystal structure, previously described, it is possible to extract the magnetic structure, as shown on figure\ref{fig:fig1}(b). In this structure, copper dimers, with an intramolecular separation of $2.628(2)$ \AA, are separated each other by a distance of $6.776(1)$\AA. These dimers are arranged in linear chains along $a$ axis, and these chains are separated each other, along $c$ axis, by a distance of $8.176(1)$ \AA, forming thus sheets of copper dimers - see figure \ref{fig:fig1}(b). Magnetically speaking, due to the much higher intra and inter chain distances, in comparison to Cu-Cu intra-dimer distance, we considered a model of isolated dimers. Considering the Cu(II) ions under consideration have a $d^9$ electronic configuration, these material is an ideal realization of a spin 1/2 dimer. The present model has one exchange parameter: $J$, related to an intra-dimer interaction corresponding to the syn-syn conformation, as can be seen at figure \ref{fig:fig1}(a). Thus, the Hamiltonian expression of this system is simply written as:
\begin{equation}
\mathcal{H}=-J \vec{S}_1\cdot \vec{S}_2-g\mu_B\vec{B}\cdot(\vec{S}_1+\vec{S}_2)
\label{eq:10}
\end{equation}
where $g$ is the isotropic Lande factor.

\begin{figure}
\centering
\includegraphics[scale=0.46]{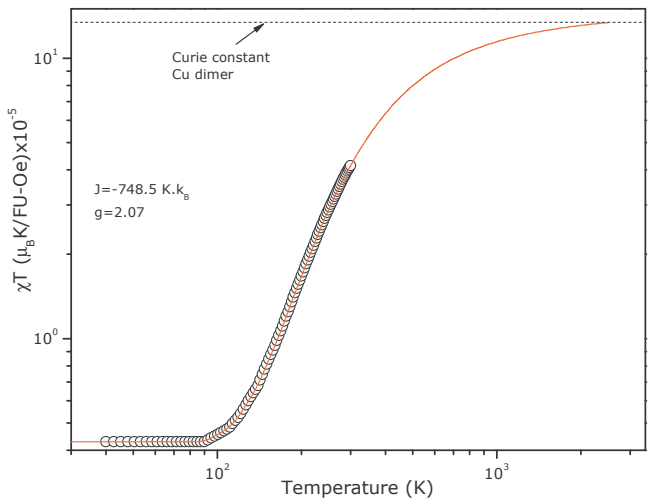}
\caption{(color online) Experimental (open circles) and theoretical (solid line) magnetic susceptibility times temperature. The red line represents the fitting of Eq.(\ref{eq:4}) to the experimental data, where an extrapolation, up to higher temperatures, was done using the optimized parameters. The dashed line indicates the Curie constant for a Cu-Cu dimer.}
\label{fig:fig2}
\end{figure}

The quantum properties of several materials have been deeply analyzing by means of thermodynamic quantities\cite{vedral5,diogo,duarte,duarte2,diogo3,souza2,mario2,vedral2,souza,vedral7,toth,ghosh,vedral8}, and magnetic susceptibility $\chi$ has been a good benchmark\cite{vedral5,diogo,duarte,duarte2,souza2,mario2,vedral2,souza,vedral8} due to the easy experimental access. For the material under consideration, the experimental $\chi T$ quantity, for the whole range of temperature considered ($<$300 K), is smaller than the Curie constant of the dimer $C=1.34\times 10^{-4}$ $\mu_B$K/FU-Oe; and it is a clear signature that at least one exchange interactions into the system is larger than the thermal energy at room temperature. As a consequence, the paramagnetic region is not reached for the temperature range in which the susceptibility is measured and therefore this experimental data cannot be described within the Curie-Weiss law. These results are shown on figure \ref{fig:fig2}. From the theoretical point of view, the magnetic susceptibility due to the Hamiltonian of equation \ref{eq:10} can be described as\cite{mario,bleaney}:
\begin{equation}
\chi (T)=\frac{2 N(g\mu_B)^2}{k_B T}\frac{1}{3+e^{-{J}/{k_B T}}}
\label{eq:4}
\end{equation}
and a fitting of the above to the experimental data (shown on figure \ref{fig:fig2}) were obtained with $J=-748.5$ K (antiferromagnetically coupled ions) and $g=2.07$. This fitting was performed using DAVE-MagProp\cite{dave}, a software that analyzes and processes magnetic data. From these fitting parameters, it was possible to extrapolate this thermodynamic quantity to higher temperatures and it is clear that the system indeed dimerizes at rather elevated temperatures, and these dimers remain isolated down to the lowest measured temperature.

To go further and analyse the quantum correlations by means of magnetic susceptibility, firstly, the spin-spin correlation function $c(T)=\langle {S}_1^{(i)}{S}_2^{(i)}\rangle$, where $i=x,y,z$, must be written as a function of the magnetic susceptibility $\chi$. For this prototype material, i.e., an ideal spin 1/2 dimer, it reads as:
\begin{equation}
c(T)=\frac{2 k_B T }{N(g\mu_B)^2}\chi (T)-1
\label{eq:5}
\end{equation}
From the above, the quantum discord can now be written as a function of the magnetic susceptibility. Quantum discord can be defined in terms of the von Neumann entropy $S(\rho_{AB})=- Tr \left[\rho_{AB}\log_2 \rho_{AB} \right]$, where $\rho_{AB}$ is the density matrix of a composite system.
The total correlation $\mathcal{I}(\rho_A:\rho_B)=S(\rho_A)+S(\rho_B)-S(\rho_{AB})$ is split into the quantum part $\mathcal{Q}$ and the classical ones $\mathcal{C}(\rho_{AB})$\cite{zurek,ma,adesso,sarandy,sarandy2,sarandy3,luo,datta,vedral}, where the classical correlation of the composite system $\rho_{AB}$ is defined as:
\begin{equation}
\mathcal{C}(\rho_{AB})=\max \left[ S(\rho_A)-\sum_{k} p_kS\left(\rho_k\right)\right]
\label{eq:10}
\end{equation} 
where the maximum is taken over all positive operator-valued measurements (POVM's) $\lbrace B_k\rbrace$ performed locally only on subsystem $B$, with the conditional state $\rho_k={(\mathbb{I}_A\otimes B_k)\rho(\mathbb{I}_A\otimes B_k)}/{p_k}$ and the probability of obtain the outcome $k$ $p_k=Tr\left[ (\mathbb{I}_A\otimes B_k)\rho(\mathbb{I}_A\otimes B_k)\right]$.

Thus, the amount of genuinely quantum correlations, called \textit{quantum discord} can be introduced as an entropic measure of quantum correlation in a quantum state, defined as the difference between the total and the classical correlation, $\mathcal{Q}(\rho_{AB})=\mathcal{I}(\rho_A:\rho_B)-\mathcal{C}(\rho_{AB})$. Thus, the entropic quantum discord depending on the magnetic susceptibility reads as:
\begin{widetext}
\begin{align}
\mathcal{Q}_E(T)=&\frac{1}{4}\lbrace \left[4-3\alpha T \chi (T)\right]\log_2 \left[4-3\alpha T\chi (T)\right]+3\alpha T\chi (T)\log_2 \left[\alpha T\chi (T)\right]\rbrace - \label{eq:6} \\\nonumber &\frac{1}{2}\lbrace\left[1+\vert \alpha T\chi (T)-1\vert\right]\log_2 \left[1+\vert \alpha T\chi (T)-1\vert\right]+ \left[1-\vert \alpha T\chi (T)-1\vert\right]\log_2 \left[1-\vert \alpha T\chi (T)-1\vert\right]\rbrace&
\end{align}
\end{widetext}
where $\alpha=2 k_B / N(g\mu_B)^2$.

On the other hand, the geometric quantum discord, based on Schatten 1-norm, is a well-defined measurement of  the amount of quantum correlations of a state in terms of its minimal distance from the set $\omega$ of classical states\cite{sarandy2,sarandy3}. Geometric quantum discord reads then as: 
\begin{eqnarray}
\mathcal{Q}_G(\rho)=\min_{\omega}\Vert\rho - \rho_c\Vert\label{eq:11}
\end{eqnarray}
where $\Vert X\Vert=Tr\left[\sqrt{X^\dagger X}\right]$ is the 1-norm, $\rho$ is a given quantum state and $\rho_c$ a closest classical-quantum state \cite{sarandy2,sarandy3}.

It is described in references \onlinecite{ciccarello,sarandy2,sarandy3,obando} for a Bell diagonal state and can now be written as a function of the magnetic susceptibility as:
\begin{eqnarray}
\mathcal{Q}_G(T)=\frac{1}{2}\left| \frac{2 k_B T }{N(g\mu_B)^2}\chi (T)-1\right|\label{eq:7}
\end{eqnarray}

Furthermore, in order to make a comparison between the quantum discord and the amount of entanglement in the system under consideration, we adopt the measure of entanglement of formation, which is oftenly used as a measurement of entanglement, defined by\cite{wootters,hill}:
\begin{equation}
\mathbb{E}=-\mathbb{E}_+-\mathbb{E}_-
\label{eq:8}
\end{equation}
where 
\begin{equation}
\mathbb{E}_\pm=\frac{1\pm\sqrt{1-\mathbb{C}^2}}{2}\log_2 \left(\frac{1\pm\sqrt{1-\mathbb{C}^2}}{2}\right)
\end{equation}
and $\mathbb{C}$ is the concurrence\cite{wootters,hill,nielsen}, that can now be written as a function of the magnetic susceptibility as:
\begin{equation}
\mathbb{C} = -\frac{1}{2}\left[ 2+3\frac{2 k_B T }{N(g\mu_B)^2} \chi (T)\right]  \qquad T< T_e\\
\label{eq:9}
\end{equation}
and zero otherwise; where $T_e\approx\vert J\vert /k_B\ln (3)\approx 0.91\vert J\vert /k_B$ is the temperature of entanglement, the maximum temperature below which there is entanglement\cite{diogo,horodecki,mario2,vedral2}.

\begin{figure}[htp]
\centering
\includegraphics[width=8.5cm]{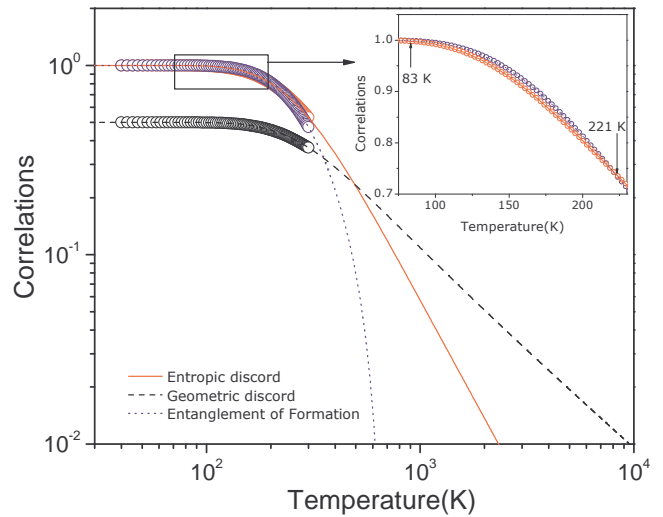}
\caption{(color online) Temperature dependence of entropic (solid red) and geometric (dashed black) quantum discords, as well as entanglement of formation (dotted blue). Due to tecnhical limitations, experimental data (open circles) are measured only up to room temperature, but the theoretical extrapolation (solid lines) goes further. Inset shows the region where the entanglement is larger than entropic quantum discord, as well as the pure state up to around $80$ K.}
\label{fig:fig3}
\end{figure}

Figure \ref{fig:fig3} shows the entropic (equation \ref{eq:6}) and geometric quantum discords (equation \ref{eq:7}), as well as the entanglement of formation (equation \ref{eq:8}), as a function of temperature, obtained from the experimental magnetic susceptibility data - open circles - and the extrapolated ones - solid lines. Note that, up to around $80$ K (above the liquid nitrogen temperature) the entropic discord and the entanglement achieve the maximum value of unity, i.e., the system is absolutely in the singlet ground state (pure state). Above this temperature, the entanglement is larger than entropic quantum discord and intercept back this curve at $221$ K. This phenomenon in which the entanglement is larger than the quantum correlations can be explained due to the fact that entanglement is a mixture of purely classical and purely quantum correlations. In addition, a direct comparison may lead to a misunderstanding, since entanglement is a different measurement of quantum correlation, as already discussed in references \onlinecite{luo,yuri,yuri2}.

From the entanglement of formation we can verify that these copper dimers are entangled up to temperatures of $T_e=681$ K when the entanglement has a sudden death. However, from the entropic quantum discord we can verify that this material can support correlated quantum states up to temperatures as high as $2320$ K. However, above $500$ K the geometric quantum discord is larger than the entropic one and it can survive up to $9540$ K, thousands of kelvins above room temperature and above the entropic quantum discord. These are the highest temperature reported on the literature wherein quantum correlation can be supported in solid-state systems. This means that the quantum correlations in this material are very stable and survives up to the temperatures in which the material exists; $523$ K\cite{paula}. Is important to emphasize that these results are only possible due to the engineered metal-carboxylate compound and its syn-syn conformation, that leads to a strong magnetic interaction ($J=-748.5$ K). Therefore, carboxylate based molecular magnets are one path toward achieving stable quantum correlations at room temperature.

In summary, our main result was to provide to the literature an engineered material with a high stability of its quantum correlations against external perturbations. We found that geometric quantum discord is significantly diferente of zero up to $9540$ K, while the entropic one shows the existence of quantum correlations up to $2320$ K; even when the entanglement is absent. Also remarkable is realization of a pure state up to $83$ K.  This prototype material has been achieved only after a successful material engineering to ensure the highest exchange interaction between spin 1/2 Cu ions into a dimeric structure. The core element to this realization is the carboxylate group, that yielded a very short (and direct) metal-to-metal extraordinarily high magnetic interaction and leads the material to be immune to decohering mechanisms. The study of this class of materials can now open a large avenue for research towards to the limits of quantum mechanics, leading to promising applications in quantum technologies.

\begin{acknowledgments}
The authors would like to thank the Brazilian funding agencies CNPq, CAPES and FAPERJ.
This work was developed within the scope of the project CICECO-Aveiro Institute of Materials, POCI-01-0145-FEDER-007679 (FCT Ref. UID /CTM /50011/2013), financed by national funds through the FCT/MEC and when appropriate co-financed by FEDER under the PT2020 Partnership Agreement.
Research at the Oak Ridge National Laboratory Spallation Neutron Source and Center for Nanophase Materials Science was sponsored by the Scientific User Facilities Division, Office of Basic Energy Sciences, of the U.S. Department of Energy. 
\end{acknowledgments}

Electronic Supplementary Information (ESI) available: CCDC 1428140 for compound $[Cu2(\mu-HCOO)4(HCOO)2]piperazine$,contains the supplementary crystallographic data for this paper. Copy of the data can be obtained free of charge on application to CCDC, 12 Union Road, Cambridge CB2 1EZ, UK fax (+44)1223 336033, e-mail: \url{deposit@ccdc.cam.ac.uk}


\begin{thebibliography}{45}
\bibitem{paula} This result was obtained by Thermogravimetric Analysis (TGA), not shown.
\end{thebibliography}
\end{document}